\documentclass[fleqn,usenatbib]{mnras}

\usepackage[T1]{fontenc}

\DeclareRobustCommand{\VAN}[3]{#2}
\let\VANthebibliography\thebibliography
\def\thebibliography{\DeclareRobustCommand{\VAN}[3]{##3}\VANthebibliography}

\usepackage{graphicx}	
\usepackage{amsmath}	
\usepackage{comment}
\usepackage{txfonts}
\usepackage{hyperref}
\usepackage{textgreek}
\usepackage{xargs}
\usepackage{xspace}
\usepackage[dvipsnames,hyperref]{xcolor}

\usepackage{hyperref}
\hypersetup{
    colorlinks=true,
    linkcolor=Cerulean,
    citecolor=NavyBlue,
    filecolor=magenta,
    urlcolor=Violet,
    pdftitle= {GSz14_ALMA},
    pdfpagemode=FullScreen,
    }

\newcommand{\Msun}{\ensuremath{\mathrm{M}_\odot}\xspace}
\newcommand{\jwst}{\textit{JWST}\xspace}

\newcommandx{\permittedEL}[6][1=O,2=III,3=,4=,5=,6=]{\text{{#1}\,{\sc {#2}}{#3}{#4}{#5}{#6}}\xspace}
\newcommandx{\semiforbiddenEL}[6][1=O,2=III,3=,4=,5=,6=]{\text{{#1}\,{\sc {#2}}]{#3}{#4}{#5}{#6}}\xspace}
\newcommandx{\forbiddenEL}[6][1=O,2=III,3=,4=,5=,6=]{\text{[{#1}\,{\sc{#2}}]{#3}{#4}{#5}{#6}}\xspace}
\newcommand{\kms}{km s$^{-1}$}

\newcommandx{\OIIL}[1][1=3728]{\forbiddenEL[O][ii][\textlambda][#1]}

\newcommandx{\OIIIL}[1][1=5007]{\forbiddenEL[O][iii][\textlambda][#1]}
\newcommand{\OIIIall}{\forbiddenEL[O][iii][\textlambda][\textlambda][5007,][4959]}

\newcommandx{\NIIL}[1][1=6585]{\forbiddenEL[N\,][ii][\textlambda][#1]}

\newcommandx{\NeIIIL}[1][1=3869]{\forbiddenEL[Ne][iii][\textlambda][#1]}

\newcommand{\CIIIall}{\semiforbiddenEL[C][iii][\textlambda][\textlambda][1907,][1909]}

\newcommand{\CII}{\forbiddenEL[C][ii][\textlambda][158$\mu$m]}
\newcommand{\OIIIalm}{\forbiddenEL[O][iii][\textlambda][88$\mu$m]}

\newcommand{\target}{JADES-GS-z14-0\xspace}

\newcommand{\JWST}{\textit{JWST}\xspace}

\title[\target kinematics]{Tentative rotation in a galaxy at z$\sim$14 with ALMA}

\author[J. Scholtz \& E. Parlanti et al.]{\parbox[h]{\textwidth}{
J. Scholtz\thanks{E-mail: js2685@cam.ac.uk}$^{1,2}$ \thanks{These authors contributed to this work equally.}, E. Parlanti \thanks{E-mail: eleonora.parlanti@sns.it}$\dagger^{3,4}$, S. Carniani$^{3}$, M. Kohandel$^{3}$, F. Sun$^{5}$, A. L. Danhaive$^{1,2}$, R. Maiolino$^{1,2,6}$, 
S. Arribas$^{7}$, R. Bhatawdekar$^{8}$, A. J.\ Bunker$^{9}$, S. Charlot$^{10}$, F. D'Eugenio$^{1,2}$, A. Ferrara$^{3}$, Z. Ji$^{11}$, Gareth C. Jones$^{1,2}$, P. Rinaldi $^{11}$, B. Robertson$^{12}$, A. Pallottini$^{3,13}$,  I. Shivaei$^{7}$, Y. Sun $^{11}$, S. Tacchella$^{1,2}$, H. \"Ubler$^{4}$, G. Venturi$^{3}$
}\vspace{0.4cm}
\\
$^{1}$ Kavli Institute for Cosmology, University of Cambridge, Madingley Road, Cambridge, 
CB3 0HA, UK\\
$^{2}$ Cavendish Laboratory, University of Cambridge, 19 JJ Thomson Avenue, Cambridge CB3 0HE, UK\\
$^{3}$ Scuola Normale Superiore, Piazza dei Cavalieri 7, I-56126 Pisa, Italy\\
$^{4}$ Max-Planck-Institut f\"ur extraterrestrische Physik (MPE), Gie{\ss}enbachstra{\ss}e 1, 85748 Garching, Germany\\
$^{5}$ Center for Astrophysics $|$ Harvard \& Smithsonian, 60 Garden St., Cambridge MA 02138 USA\\
$^{6}$ Department of Physics and Astronomy, University College London, Gower Street, London WC1E 6BT, UK\\
$^{7}$ Centro de Astrobiolog\'ia (CAB), CSIC–INTA, Cra. de Ajalvir Km.~4, 28850- Torrej\'on de Ardoz, Madrid, Spain\\
$^{8}$ European Space Agency (ESA), European Space Astronomy Centre (ESAC), Camino Bajo del Castillo s/n, 28692 Villanueva de la Cañada, Madrid, Spain \\
$^{9}$ University of Oxford, Department of Physics, Denys Wilkinson Building, Keble Road, Oxford OX13RH, United Kingdom\\
$^{10}$ Sorbonne Universit\'e, CNRS, UMR 7095, Institut d'Astrophysique de Paris, 98 bis bd Arago, 75014 Paris, France\\
$^{11}$ Steward Observatory, University of Arizona, 933 N. Cherry Avenue, Tucson, AZ 85721, USA\\
$^{12}$ Department of Astronomy and Astrophysics University of California, Santa Cruz, 1156 High Street, Santa Cruz CA 96054, USA\\
$^{13}$ Dipartimento di Fisica ``Enrico Fermi'', Universit\'{a} di Pisa, Largo Bruno Pontecorvo 3, Pisa I-56127, Italy \\
}

\date{Accepted XXX. Received YYY; in original form ZZZ}

\pubyear{2024}

\begin{document}
\label{firstpage}
\pagerange{\pageref{firstpage}--\pageref{lastpage}}
\maketitle

\begin{abstract}
We re-analysed ALMA observations of the \OIIIalm emission line in \target, so one of the most distant spectroscopically confirmed galaxy at z=14.18. Our analysis shows a tentative detection of a velocity gradient of \OIIIalm using three independent tests: 1) construction of moment maps; 2) extraction of integrated spectra from a grid of apertures; and 3) spectro-astrometry in both the image and \textit{uv} planes, confirming the presence of the velocity gradient at 3$\sigma$ significance. We performed kinematical fitting using the KinMS code and estimated a dynamical mass of log$_{10}$(M$_{\rm dyn}$/$\rm M_\odot$)= 9.4$^{+0.8}_{-0.4}$, with the bulk of the uncertainties due to the degeneracy between dynamical mass and inclination. We measure an upper limit on the velocity dispersion ($\sigma_{v}$) of $<40~$\kms~which results in an estimate of V$_{\rm rot}/\sigma>$ 2.5. This result, if confirmed with higher-resolution observations, would imply that kinematically cold discs are already in place at $z\sim14$. Comparison with mock observations from the SERRA cosmological simulations confirms that even low-resolution observations are capable of detecting a velocity gradient in $z>10$  galaxies as compact as \target. This work shows that deeper ALMA or \jwst/NIRSpec IFS observations with high spatial resolution will be able to estimate an accurate dynamical mass for \target, providing an upper limit to the stellar mass of this over-luminous galaxy.

\end{abstract}

\begin{keywords}
galaxies; kinematics \& dynamics --- galaxies: evolution;
\end{keywords}



\section{Introduction}

With the launch of the James Webb Space Telescope (\JWST), we are now able to observe the rest-frame optical and UV emission from galaxies and their interstellar medium (ISM) up to redshift $\sim$ 14 \citep{Cameron23,Curtis-lake23, Harikane23,Larson23,Isobe23,Hsiao23,Robertson23, Abdurrouf24, Carniani24_gsz14_jwst, Harikane24, Sanders24,Tacchella23,Tacchella24, Vikaeus24}, revealing a large population of bright, metal-poor galaxies in the early Universe. Remarkably, insight into the characteristics of the most distant galaxies can also be gained through interferometric observations, as illustrated by the detection of line emission from what is currently the most distant galaxy known (ALMA; \citealt{Carniani24_gsz14_ALMA, Schouws24}).

Before the launch of \JWST, the main avenue to study galaxies properties of high-redshift galaxies was with HST and ground-based 8-10 m observatories to study the rest-frame UV properties. Mm/sub-mm observatories such as ALMA and NOEMA investigated these high-z galaxies through \CII\ and \OIIIalm emission lines and in a few cases far-IR continuum emission. These observations revealed large dust reservoirs by z$\sim$8 \citep[][]{Tamura19, Bakx21, Inami22, Sommovigo22, Witstok22}, but also ISM properties such as metallicity and ionisation parameter \citep[][]{Spilker22, Witstok22, Litke23, Killi23}.

Far infra-red (FIR) observations of high-z galaxies also showed early rotating discs \citep[e.g.][]{Smit18, Neelman20, Rizzo20, Fraternali21, Lelli21, Rizzo21, Parlanti23_alm, Rowland24}. However, in some cases, \JWST observations have resolved these apparent discs into close-separation major mergers \citep[e.g.][]{Fraternali21, Jones24, Lamperti24,Scholtz24_COS30,Parlanti24_HZ4, Ubler24b}, although there is still debate over different kinematic signatures in cold and ionised gas phases. Theoretical models predict that high redshift galaxies should be more turbulent, due to the increased merger rates \citep[][]{duncan19,Duan:2024}, violent disc instabilities driven by the accretion of gas  \citep[][]{dekel_cold_2009, krumholz_discs_2018}, and intense star formation feedback \citep[][]{Orr2020}. However, different cosmological simulations find different results, with some showing the presence of turbulent discs \citep[][]{Pillepich:2019}, 
while others predict that early galaxies should already have formed a cold rotating disc \citep[][]{Kohandel24}. This still leaves major questions about the assembly of discs at the epoch of reionisation and beyond, including which gas phase best traces the galaxy kinematics/rotation.

The most distant spectroscopically confirmed galaxy is JADES-GS-53.08294-27.85563 \citep[more commonly known as \target;][]{robertson_JOF_LF_2024} at z = 14.1796, originally identified via NIRCam photometry \citep[][]{hainline_zgtr10_jades_2024, robertson_JOF_LF_2024} and later spectroscopically confirmed by NIRSpec \citep[][]{Carniani24_gsz14_jwst}, and ALMA \citep[][]{Carniani24_gsz14_ALMA, Schouws24}. The combination of ALMA and \jwst observations revealed a compact galaxy with a half-light radius of 260 $\pm$ 20 pc
with metallicity of 0.1-0.2 Z$_{\odot}$, i.e. a significantly enriched ISM for z$\sim$14 galaxy \citep[][]{Carniani24_gsz14_jwst}. The galaxy is also detected at 7.7\,\micron\ with MIRI with strong excess to NIRCam photometry, indicating the presence of \OIIIall\ \citep{Helton_2024_z14}. Furthermore, \citet{Carniani24_gsz14_ALMA} and \citet{Schouws24} detected \OIIIalm in this galaxy with ALMA, measuring the dynamical mass from the velocity dispersion of the \OIIIalm (log$_{10}$(M$_{\rm dyn}$/\Msun) = 9.0$\pm$0.2), comparable to the galaxy's stellar mass (log$_{10}$(M$_{*}$/\Msun) = 8.7$^{+0.7}_{-0.2}$). Under standard assumptions about stellar-population modelling (e.g., \citealp{Chabrier03} initial mass function; \citealt{Carniani24_gsz14_ALMA, Helton_2024_z14}), the measured dynamical mass leaves very little mass budget for both gas and  stars -- the estimated gas fraction is only 10-30 \%, making this galaxy gas-poor. This would be a result of either rapid gas consumption through previous star-formation episodes or fast gas outflows ejecting the gas reservoir \citep[][]{Tacchella16, Dekel2023, Ferrara_2024, Ferrara_2024b}.

However, the dynamical mass estimates using the velocity dispersion of a line are notoriously uncertain, hence an accurate measurement of the dynamical mass with dynamical modelling is needed \citep[][]{Kohandel19}. 
In this Letter, we investigate the evidence for a velocity gradient using the ALMA observations of \OIIIalm, to improve the measurement of the dynamical mass of \target and investigate the assembly of rotating discs in the early Universe. 

Throughout this work, we adopt a flat $\Lambda$CDM cosmology: H$_0$: 67.4 km s$^{-1}$ Mpc$^{-1}$, $\Omega_\mathrm{m}$ = 0.315, and $\Omega_\Lambda$ = 0.685 \citep{2020A&A...641A...6P}. With this cosmology, $1''$ corresponds to  3.3\,kpc at $z=14.18$.

\section{Observations and data reduction}
\label{sec:obs}

In this work, we use ALMA band 6 observations of \target as part of 2023.A.00037.S programme (PI: Schouws), which used two spectral configurations to target the \OIIIalm emission line. The total on-source integration time is 2.82 hours per spectral configuration. We used the calibrated visibilities downloaded from the ALMA archive, only selecting spectral configuration covering the 
detected \OIIIalm emission line at 223.524 GHz, same data reduction as \citet[][]{Carniani24_gsz14_ALMA} 

In order to image the calibrated visibilities we used the Common Astronomy Software Applications package (CASA; \citealt{McMullin07,casa22}), specifically the task \verb|tclean|. \citet{Carniani24_gsz14_ALMA} and \citet{Schouws24} detected the emission line with 6.7$\sigma$. To optimize the spatial resolution of our observations while maintaining sufficient SNR, we adopted Briggs weighting with a robust parameter of 0.5 and hogbom deconvolver. We opted for a pixel scale of 0.1\arcsec, and a channel width of 10~\kms \xspace to create the final datacubes. We cleaned the images down to 3$\sigma$ level (rms= 0.1 mJy/beam). The final data cube has a resolution with a beam of 0.6\arcsec\ $\times$ 0.8\arcsec, compared to the deconvolved rest-frame UV half-light radius from NIRCam imaging of 0.079\arcsec and the natural weighting resolutin of 1.09\arcsec$\times$ 0.82\arcsec used in \citet{Carniani24_gsz14_ALMA}\citet[][]{Schouws24}.

\section{Detection and modelling of velocity gradient}

\subsection{Velocity gradient in \OIIIalm}\label{sec:gradient}

\begin{figure*}
        \centering
	\includegraphics[width=0.99\textwidth]{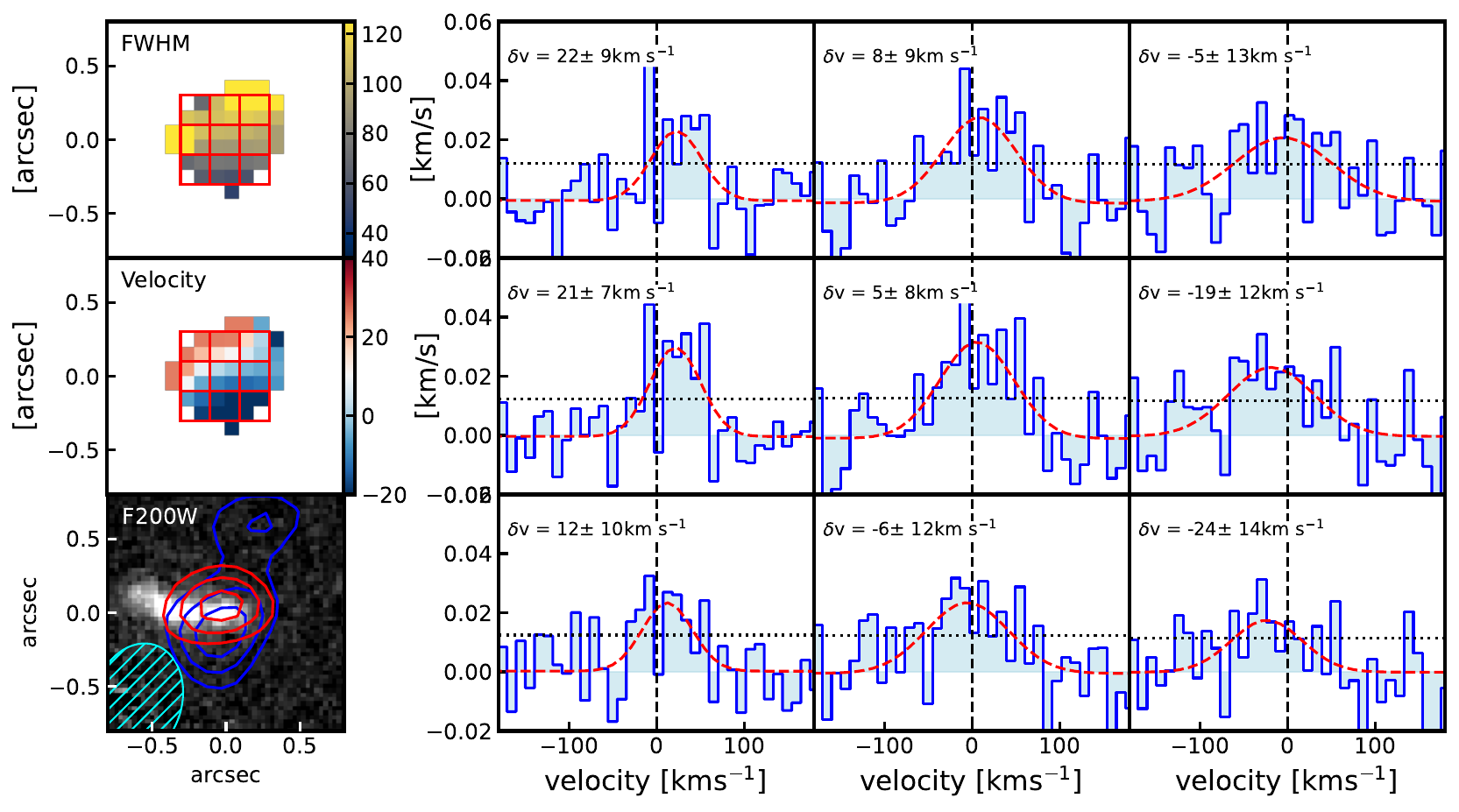}
    \caption{ Tentative detection of the velocity gradient in \target.
    Left top and middle panels: FWHM and velocity maps of the \OIIIalm emission line from Gaussian fitting. We used a 3$\sigma$ cutoff to create these maps. The red squares indicate the apertures used to extract spectra on the right. Bottom left: NIRCam F200W image tracing the rest-frame UV emission. The red and blue contours show moment-0 maps extracted from [$-50$, 0] \kms and [0, 50] \kms. The cyan-hatched ellipse shows the ALMA beam size. We observe 0.3$\pm$0.06 \arcsec\. offset between the red and blue centroids. Right panels: \OIIIalm extracted from the square apertures indicated on velocity and FWHM maps. The data and the best-fit model are shown as blue and red lines, while the black dotted line indicates the flux uncertainties. The extracted spectra confirm the derived velocity maps.}
    \label{fig:maps}
\end{figure*}

To investigate the presence of any velocity gradient traced by the \OIIIalm emission line, we fitted each pixel of the data cube within a radius of $<$0.5\arcsec\ with a single Gaussian model, with centroid, velocity width and amplitude as free parameters. We fitted this model using python's \texttt{lmfit}. To construct the final map we opted for a SNR cutoff of 3. We show the final velocity and FWHM maps of \OIIIalm in the top and middle left panels of Fig. \ref{fig:maps}. The velocity map shows a velocity gradient in the northeast direction with velocities from +40 to $-20$ \kms. We verified this velocity gradient using three separate methods: 1) extracting integrated spectra from grid of apertures; 2) performing spectro-astrometry in the image plane; and 3) performing spectro-astrometry in the uv-plane.

To confirm the velocity gradient, we extracted the regional spectra from 3$\times$3 square grid centred on the \target. Each region has a size of 0.3\arcsec$\times$0.3\arcsec\ and we show these spectra on the right side of Fig. \ref{fig:maps}. We fitted the extracted spectra using a single Gaussian profile to determine the velocity centroid of the emission line profile. The extracted spectra confirm the presence of the velocity gradient in \target along the north-east direction with maximum velocity difference between two regions of 42$\pm$14 \kms \xspace, which is significant at 3.0$\sigma$.

We further confirmed this velocity gradient using spectro-astrometry in both image and \textit{uv}-plane. For the image plane analysis, we created two moment-0 maps of the \OIIIalm emission line in the range [0, 50] \kms \xspace and [$-50$, 0] \kms \xspace to map of the redshifted and blueshifted parts of the emission line, respectively. These are shown as red and blue contours, respectively, on the moment-0 map in the bottom left panel of Fig. \ref{fig:maps}. We measure the spatial offset between the red and blue centroids of 0.30$\pm$0.06\arcsec, corresponding to 0.99$\pm$0.20 kpc.

\begin{figure}
        \centering
	\includegraphics[width=0.99\columnwidth]{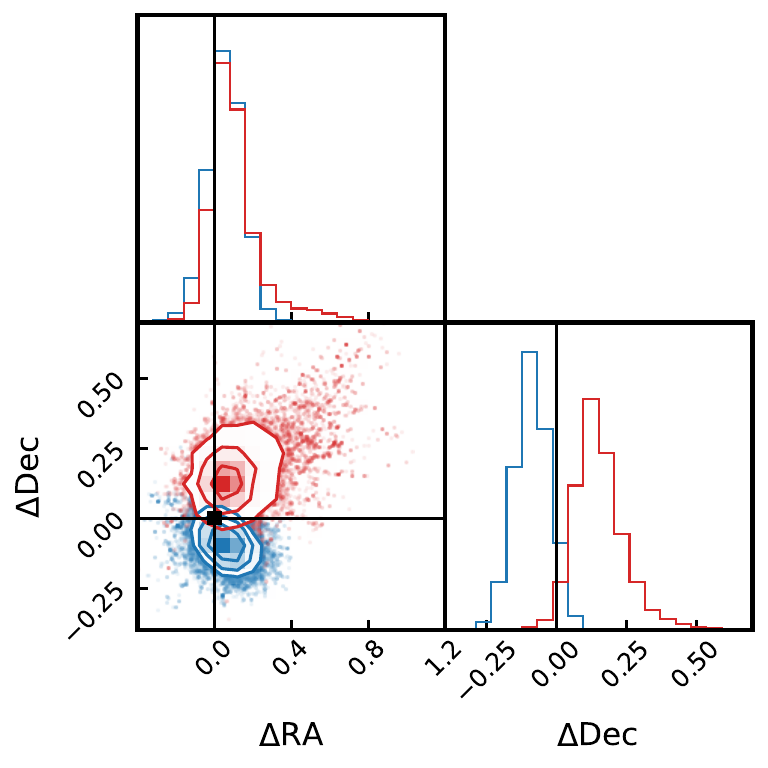}
    \caption{ Spectro-astrometric analysis in the uv-plane. We show the posterior distribution of the spatial offset of the red and blue sides of the emission line from the centre of \target ([0, 50] \kms \xspace and [$-50$, 0] \kms). We see a spatial offset between the red and blue parts of the emission line, confirming the presence of the velocity gradient.}
    \label{fig:uv}
\end{figure}

We adopted a spectro-astrometry test in the \textit{uv}-plane by performing a 2D Gaussian fitting to the \textit{uv}-visibilities. We first extracted the \textit{uv}-visibilities from the measurement set using spectral channels 66--70 for the blue side and channels 70--74 for the red side of the emission line, which corresponds to the [0, 50] \kms \xspace and [$-50$, 0] \kms \xspace channel maps. We fitted the model using MCMC fitting routine (\texttt{emcee}; \citealt{foreman-mackey+2013}) and we plot the final posterior distribution of the offsets between the location of the red and blue side of the emission line and the centre of the galaxy in Fig. \ref{fig:uv}. The final estimated distance between the red and blue sides is 0.28$\pm$0.10\arcsec\ (0.92$\pm$0.33 kpc), clearly confirming the velocity gradient observed in the image plane. Furthermore, we split the observations into two exposures (1.4 hours on source per exposure) and confirmed our findings in both exposures. 
Overall, all three approaches show tentative evidence for a velocity gradient in \target.

\subsection{Modelling of the velocity gradient}\label{sec:kinfit}

In this section, we model the tentative velocity gradient detected in the previous section. Interpreting velocity gradients in low-resolution observations is challenging as mergers are often misinterpreted as rotating discs (e.g., 
\citealt{Rizzo22, Kohandel_2020, Scholtz24_COS30, Parlanti24_HZ4}). However, in the NIRCam images tracing the rest-frame UV (see Fig. \ref{fig:maps}), we do not see evidence of companions, double nuclei, tidal tails or any other complex structure usually associated with merger events. The presence of an outflow causing the observed gradient is also excluded based on the low velocity dispersion of the observations. Within the SNR of our observations, we do not either see clear traces of streams of inflowing gas that could lead to kinematic distortions and dynamical instabilities like those observed in other high-z systems (e.g. SPT0311-58; \citealt{Arribas23}). The measured offset of the red and blue centroids of \OIIIalm\ spectrum is also well above the diffraction limit of JWST at 3\,\micron\ (0.1\arcsec), implying that any companion or tidal tails above this spatial scale would have been detected in the NIRCam images. Therefore, we will model this data by assuming it is a rotating disc. 

We use the same method as described in \citet[][]{Parlanti23_alm}, using the publicly available \texttt{Python} library \textsc{KinMS} \citep[][]{Davis20}. Here we briefly describe the procedure. This method creates a mock data cube of a rotating disc based on the input parameters, convolves it to match the spatial resolution of the observation to account for the beam-smearing effect, and then creates the model moment maps that we can directly compare with a set of observed ones.
We set up  KinMS to simulate the ALMA observations, with a spectral resolution of 10 \kms\ and angular resolution set to the beam size of our observations (0.6\arcsec$\times$0.8\arcsec\ with an angle of 89\textdegree). To model the galaxy we assume that the matter is distributed in a thin exponential disc \citep{Freeman1970} with an intrinsic constant velocity dispersion profile. We set a uniform prior on the $\sigma_{v}$ between 0--200 \kms \xspace.
Using this setup we generate a mock cube and moment maps which are compared to the observed ones. The best-fit parameters are estimated using the package \textsc{emcee} \citep[][]{foreman-mackey+2013}.  We used uniform priors on the inclination between 5\textdegree and 85\textdegree, and uniform priors on position angle between 0\textdegree and 180\textdegree. The dynamical mass was left free to vary with log uniform priors between $10^8$ and $10^{11}$ $M_{\odot}$ and we assumed uniform priors on the velocity dispersion between 5 and 200 \kms. We masked the pixels in the Moment-2 map with FWHM$>$150 \kms \xspace as they are dominated by noise on the outskirts of the galaxy.

We fitted simultaneously the velocity and velocity dispersion maps derived in \S~\ref{sec:gradient} to estimate the inclination of the disc, dynamical mass, position angle, and intrinsic velocity dispersion (i.e. deconvolved by the instrumental spatial resolution), while we fixed the disc effective radius to the one found in \citet{Carniani24_gsz14_jwst} from NIRCam imaging of 260 pc (0.079 arcsec). We note that the stellar and ionised gas sizes were found to be comparable in the IFU observations \citep[e.g.][]{ForsterSch18b, Scholtz18}.

\begin{figure}
        \centering
	\includegraphics[width=0.99\columnwidth]{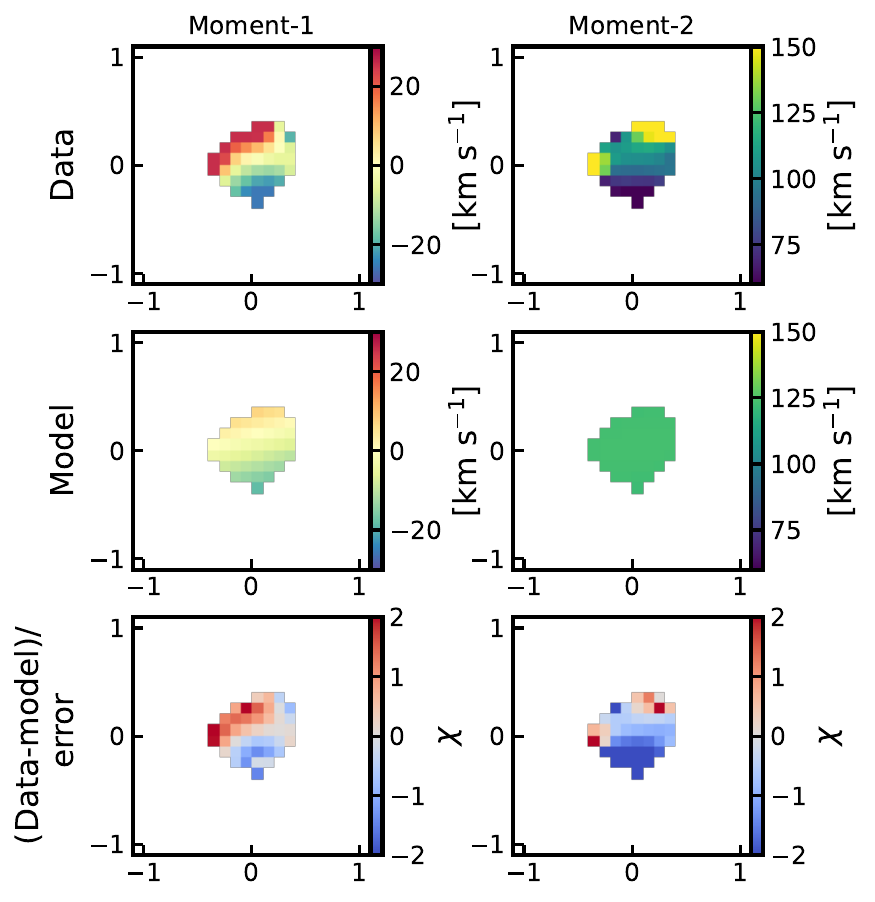}
    \caption{Moment maps created from the observed data (top row), maps from the best fit kinematical model of KinMS (centre row), and the normalised residuals (bottom row).  The left and right columns show moment-1 and moment-2 maps.
    }
    \label{fig:Kin_mod}
\end{figure}

We show the comparison of the data and the fitted model in Fig. \ref{fig:Kin_mod}. We note that the model is not fully able to reproduce the observed velocity gradient,  due to the low value of the observed velocity dispersion. To reproduce the observed velocity gradient, the beam-smearing effect does increase the observed velocity dispersion. Modelling only the velocity gradient would improve the fit of the velocity map and result in a higher dynamical mass, but would overestimate the observed velocity dispersion map. In this work, we used the approach of fitting the maps simultaneously. With low-resolution data, the beam-smearing effect causes the observed velocity dispersion to be affected by the underlying velocity gradient, adding little constraint to the overall fit. Higher-resolution observations would decrease the level of beam-smearing, allowing for more accurate modelling.

\begin{figure}
        \centering
	\includegraphics[width=0.99\columnwidth]{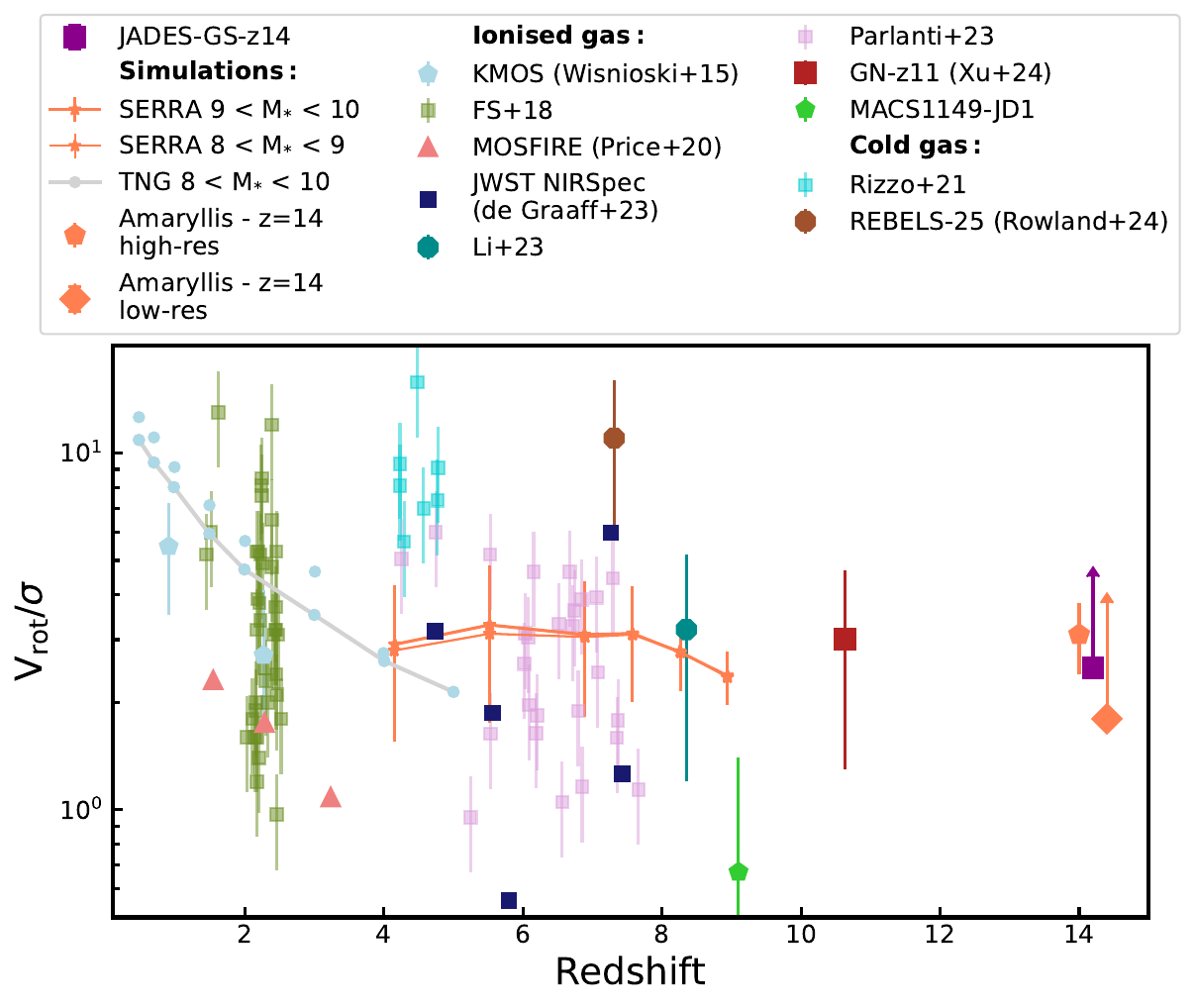}
    \caption{V$_{\rm rot}$/$\sigma_v$ versus redshift for observations and simulations. As we are unable to constrain the velocity dispersion of \target and in the low-resolution mock observations of Amaryllis from SERRA simulations, we only quote a lower limit on V$_{\rm rot}$/$\sigma_{v}$ for these. We compare \target to a compilation from the literature: \jwst/NIRSpec, Illustris TNG \citep[][]{Pillepich:2019}, SERRA \citep[][]{Kohandel24}, ground-based IFU \citep[][]{Wisnioski15, forster_schreiber_sinszc-sinf_2018}, MOSFIRE \citep[][]{price_mosdef_2019}, GN-z11 \citep[][]{Xu24_gnz11kin}, ALMA z$\sim$6 \citep[][]{Parlanti23_alm}, NIRSpec/MSA \citep[][]{degraaff24}, ALMA z$\sim$4 \citep[][]{Rizzo21}, REBELS-25 \citep[][]{Rowland24}, MACS1149-JD1 \citep[][]{Tokuoka22} and  MACS0416-Y3 \citep[][]{Li+23}.}
    \label{fig:vsigz}
\end{figure}

Due to the low resolution of the data, we cannot simultaneously obtain tight constraints on the inclination and the dynamical mass of the system as they are degenerate (in low-resolution observations, models with constant M$_{\rm dyn}$ $\sin^2{(inc)}$ are similar, see Figure \ref{fig:corner}). Due to the form of the degeneracy, even when the inclination is unconstrained, we are able to estimate, with large uncertainties, the dynamical mass of the system, obtaining a value of log(M$_{\rm dyn}/$\Msun) = 9.4$^{+0.8}_{-0.4}$.
Assuming that the mass is distributed as an exponential disc with the fixed effective radius of 260 pc and the dynamical mass obtained by the kinematic modelling, we computed the rotational velocity as the maximum velocity of the rotating disc occurring at 2.2 $r_d$ $\sim$ 340 pc. 
With these assumptions, we obtain an estimate of the rotational velocity of V$_{\rm rot} = 164_{-60}^{+248}$ \kms.
Interestingly we find a 3$\sigma$ upper limit on the intrinsic velocity dispersion of $< 40$ \kms\ (3$\sigma$ upper limit). This would result in an estimation of V$_{\rm rot}/\sigma_{v} > 2.5$, which would imply tentative evidence for a kinematically cold disc in the first 300 Myr of the Universe's lifetime. We compare our measurement of V$_{\rm rot}/\sigma_{v}$ with simulations and observations compiled from the literature by placing this estimate on a plot of V$_{\rm rot}$/$\sigma_{v}$ versus redshift (see Fig. \ref{fig:vsigz}). We find that our measured tentative value for \target aligns with simulation predictions. It follows the evolutionary trend of
M$_{*}>10^{9}$ \Msun systems at z$>$6. If confirmed, this would be the most distant dynamically settled disc structure observed to date.

Our measurement of M$_{\rm dyn}$ is consistent within 1$\sigma$ with the values measured by \citet{Carniani24_gsz14_ALMA} ($\log$(M$_{\rm dyn}/\Msun$) = 9.0$\pm0.2$) and \citet{Schouws24}  ($\log$(M$_{\rm dyn}/\Msun$) = 9.0$\pm0.3 \sin(i)^{2}$)from the velocity dispersion of the \OIIIalm, but our errorbars fully capture the additional uncertainty due to the mass-inclination degeneracy.
We derive a gas fraction of this galaxy of 60$\pm$20\%, consistent with the value derived by previous analysis of the \OIIIalm observations of 36\% \citep[][]{Schouws24, Carniani24_gsz14_ALMA} and $<$70\% \citep[][]{Schouws25}.
However, for better constraints on the kinematics of \target, we require higher sensitivity and resolution observations from ALMA or \jwst/NIRSpec-IFS (see the results of our simulations in \S \ref{sec:SERRA}).

We also fitted the 1D velocity and velocity dispersion profiles, and the moment maps by using the dynamical fitting code DysmalPy \citep[][]{Price21, Lee24}, which has proven to be a reliable method to recover the kinematical properties of datasets with low S/N and low spatial resolution. With this method, we obtain comparable results (within 1$\sigma$) for the dynamical mass and the velocity dispersion.

\section{Comparison to SERRA simulations}\label{sec:SERRA}

We compare our results against mock \OIIIalm data generated from the SERRA suite of zoom-in, high-resolution cosmological simulations \citep{pallottini_serra_2022}. These simulations achieve a mass resolution of $1.2\times 10^{4} $ \Msun, $\sim$25 pc at z = 7.7, and incorporate non-equilibrium chemistry with on-the-fly radiative transfer. Their outputs are subsequently post-processed to generate FIR 
emission-lines \citep{pallottini_2019} and hyperspectral datacubes \citep{Kohandel_2020}, well-suited for producing realistic mock observations of the early Universe. In particular, we use the simulated galaxy ``Amaryllis'', identified by \citet{Kohandel23} as the brightest galaxy ($\rm{log}(L_{[OIII]_{88\mu m}}/L_\odot)\approx 8.4$) in SERRA at $11<z<14$. Amaryllis has a stellar mass $\rm{log}(M_\star/M_\odot) \approx 8.8$ and $\rm{SFR}_{10\,\rm{Myr}}=18\,M_\odot\,\mathrm{yr}^{-1}$, making it a close analogue to \target. To generate synthetic ALMA observations, we select a viewing angle of 45\textdegree, at which the FWHM of the \OIIIalm emission line closely matched that of \target. The effective radius of Amaryllis is $\sim$ 180 pc, while \target has a size of 260 pc. To match this scale, we rescale the mock data by a factor of 1.5 to improve the comparison between the mock observations and \target. 

\begin{figure*}
        \centering
	\includegraphics[width=0.8\paperwidth]{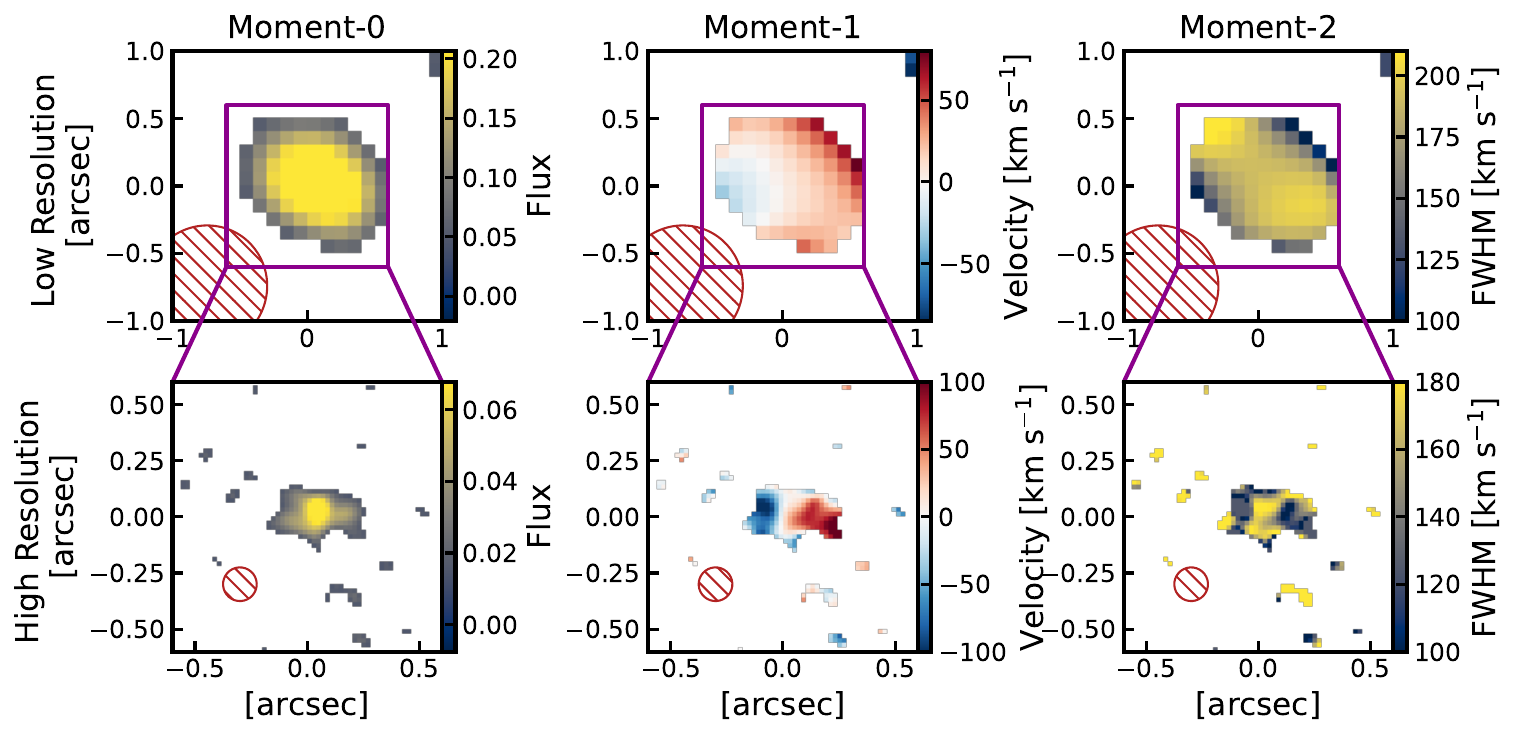}
    \caption{Mock observations of \target analogue in the SERRA simulations - "Amaryllis". The top row shows flux, velocity and FWHM maps of mock low-resolution  ALMA observations (0.9\arcsec), while the bottom row shows mock high-resolution observations (0.15\arcsec). The simulations show that a rotating galaxy shows a small velocity gradient ($<40$ \kms) in the low-resolution data, similar to our observations. The simulations show that a rotating (thin) disc galaxy with a size and mass as \target, and observed under similar conditions, should show a similar velocity gradient to the one obtained. The increased FWHM in the center of high-resolution observations can be due to the beam smearing of the rotation.}
    \label{fig:Serra}
\end{figure*}

We used CASA's \texttt{simobserve} with an array configuration of C5 and C7 corresponding to a resolution of 0.8\arcsec and 0.15\arcsec, respectively. We use integration times of 2.85 and 10.0 hours for the C5 and c7 array configurations, respectively, to match the SNR of our current observations and of potential future deeper, higher-resolution ALMA and NIRSpec/IFS observations.. We show the moment maps of Amaryllis (flux, velocity and FWHM) in Fig. \ref{fig:Serra} for the low and high resolution observations.

The mock observations from the SERRA simulations in Fig. \ref{fig:Serra} show that a velocity gradient in a rotating galaxy at z$>$10 like \target can be detected even in low-resolution observations (see top row of Fig. \ref{fig:Serra}). We modelled the mock observations using the same modelling setup described in \S~\ref{sec:kinfit}. Despite detecting the velocity gradient in the low resolution mock observations of Amaryllis simulations, we encountered similar degeneracies between inclination and dynamical mass as for the ALMA observations of \target and we can only estimate an upper limit to the velocity dispersion, therefore we are only able to estimate a lower limit on V$_{\rm rot}/\sigma_{v}$> 1.8. However, using high spatial resolution observations ($\sim$0.15\arcsec), which are achievable by both \jwst/NIRSpec and ALMA observations, it would be possible to resolve these high-z targets with more than three independent resolution elements (see beam size in Fig. \ref{fig:Serra}), which is the required condition to properly constrain a galaxy kinematics and distinguish rotating disks from close mergers \citep[][]{Rizzo22}. Indeed we tested a merger scenario of Amaryllis from the SERRA simulations in the Appendix \ref{s.merger}, showing that we are capable of distinguishing between a merger and disk rotation scenario with high resolution observations. For \jwst/NIRSpec-IFS observation, we would need to target the \CIIIall emission line doublet as it is the only detected emission line in the current spectrum, requiring the on source exposure time of over $\sim$50 hours. However, given the PSF size of 0.15\arcsec, these observations would able to resolve the blue and red side of the emission, as measured in Fig. \ref{sec:gradient}.  With such deeper high-resolution observations, it would be possible to break the degeneracy between the dynamical mass and inclination (see Fig. \ref{fig:corner} in Appendix, right panel) with a measured value of V$_{\rm rot}/\sigma_{v}= 3.1\pm0.7$ and an uncertainty on the dynamical mass of 0.2 dex. 
This clearly shows that a follow-up with deep JWST/NIRSpec-IFS or ALMA observations is essential to constrain the kinematics of \target.

\section{Conclusions}\label{sec:conclusions}

In this letter, we re-examined the ALMA observations of \OIIIalm emission line in \target at z$\sim$14.2, to constrain its kinematic properties. We created velocity maps, by fitting a Gaussian profile to \OIIIalm emission line, and we found evidence for a velocity gradient in the data with a maximum velocity difference of 50 \kms\ in the north-south direction. We confirmed the presence of the velocity gradient by 1) extracting regional spectra from 9 regions (see Fig. \ref{fig:maps}); and 2) performing  spectro-astrometry in the image and \textit{uv} planes, which found a spatial offset of  0.28$\pm$0.10\arcsec\ between the redhifted and blueshifted parts of the emission line (see Figs. \ref{fig:maps} and \ref{fig:uv}). Using these approaches we found similar velocity gradients.

Due to the lack of merger signatures and/or streams of gas associated to outflows and accretion of gas in the NIRCam imaging, we modelled the galaxy kinematics assuming a thin rotating disc of size 260 pc. We measured a dynamical mass of log(M$_{\rm dyn}$/\Msun) = 9.4$^{+0.8}_{-0.4}$. Our large uncertainties reflect the inability of the data with low spatial resolution to break the degeneracy between dynamical mass and inclination. However, our model fully captures this additional uncertainty, which is not accounted for by the virial estimator. We estimated a lower limit on the V$_{\rm rot}/\sigma_v$ of $>$2.5, which would indicate a settled dynamically cold at these high redshifts if confirmed. 

We compared our observational results with mock observations from the SERRA cosmological simulations. We created mock ALMA \OIIIalm observations of an analogue of \target called ``Amaryllis'', with both low and high spatial resolution setup (0.8\arcsec\ and 0.15\arcsec). We show that the tentative detection of the velocity gradient is possible even in the low-resolution data (see the top row of Fig. \ref{fig:Serra}). However, in order to break the degeneracy between dynamical mass and inclination, high spatial resolution observations ($<$0.2\arcsec) are required. These high spatial resolution observations could confirm the high V$_{\rm rot}/\sigma_{v}$ in the galaxy and constrain the dynamical mass within 0.23 dex. 

Our tentative detection of this velocity gradient suggests that the formation of early cold rotating structures in the Universe happened much sooner than previously anticipated in both observations and simulations. 
The discovery of a rotation-dominated disc just 300 Myr after the Big Bang would drastically change our view of early galaxies, and pose new challenges to the cosmological simulations, while putting constraints on feedback models. However, due to the limited resolution of the current observation we can only obtain a tentative detection of the presence of an early rotating disc. Further high spectral and higher spatial resolution deep observations with ALMA and \jwst IFU are essential in order to resolve the kinematics of \target and investigate the formation of discs in the early Universe.

\section*{Acknowledgements}
This paper makes use of the following ALMA data: ADS/JAO.ALMA\#2023.A.00037.S ALMA is a partnership of ESO (representing its member states), NSF (USA) and NINS (Japan), together with NRC (Canada), MOST and ASIAA (Taiwan), and KASI (Republic of Korea), in cooperation with the Republic of Chile. The Joint ALMA Observatory is operated by ESO, AUI/NRAO and NAOJ. The National Radio Astronomy Observatory is a facility of the National Science Foundation operated under a cooperative agreement by Associated Universities, Inc.
JS, FDE, RM and GCJ acknowledge support by the Science and Technology Facilities Council (STFC), ERC Advanced Grant 695671 ``QUENCH" and the
UKRI Frontier Research grant RISEandFALL. RM also acknowledges funding from a research professorship from the Royal Society.
S.C, EP and GV acknowledge support from the European Union (ERC, WINGS,101040227)
A.L.D. thanks the University of Cambridge Harding Distinguished Postgraduate Scholars Programme and Technology Facilities Council (STFC) Center for Doctoral Training (CDT) in Data intensive science at the University of Cambridge (STFC grant number 2742605) for a PhD studentship.
SA acknowledges grant PID2021-127718NB-I00 funded by the Spanish Ministry of Science and Innovation/State Agency of Research (MICIN/AEI/ 10.13039/501100011033)
AJB acknowledge funding from the "FirstGalaxies" Advanced Grant from the European Research Council (ERC) under the European Union’s Horizon 2020 research and innovation programme (Grant agreement No. 789056).
ZJ acknowledges JWST/NIRCam contract to the University of Arizona, NAS5-02015.
BER acknowledges support from the NIRCam Science Team contract to the University of Arizona, NAS5-02015, and JWST Program 3215.
We acknowledge the CINECA award under the ISCRA initiative, for the availability of high performance computing resources and support from the Class B project SERRA HP10BPUZ8F (PI: Pallottini). We gratefully acknowledge the computational resources of the Center for High Performance Computing (CHPC) at SNS.
H\"U acknowledges support through the ERC Starting Grant 101164796 ``APEX''.
IL acknowledges support from PID2022-140483NB-C22 funded by AEI 10.13039/501100011033 and BDC 20221289 funded by MCIN by the Recovery, Transformation and Resilience Plan from the Spanish State, and by NextGenerationEU from the European Union through the Recovery and Resilience Facility
SA acknowledge grant PID2021-127718NB-I00 funded by the Spanish Ministry of Science and Innovation/State Agency of Research (MICIN/AEI/ 10.13039/501100011033)

\section*{Data Availability}

The datasets were derived from sources in the public domain: ALMA data from \url{https://almascience.nrao.edu/aq/?result_view=observation}.

\appendix 
\section{Posterior distribution of kinematical fitting}
\begin{figure*}
        \centering
	\includegraphics[width=0.48\textwidth]{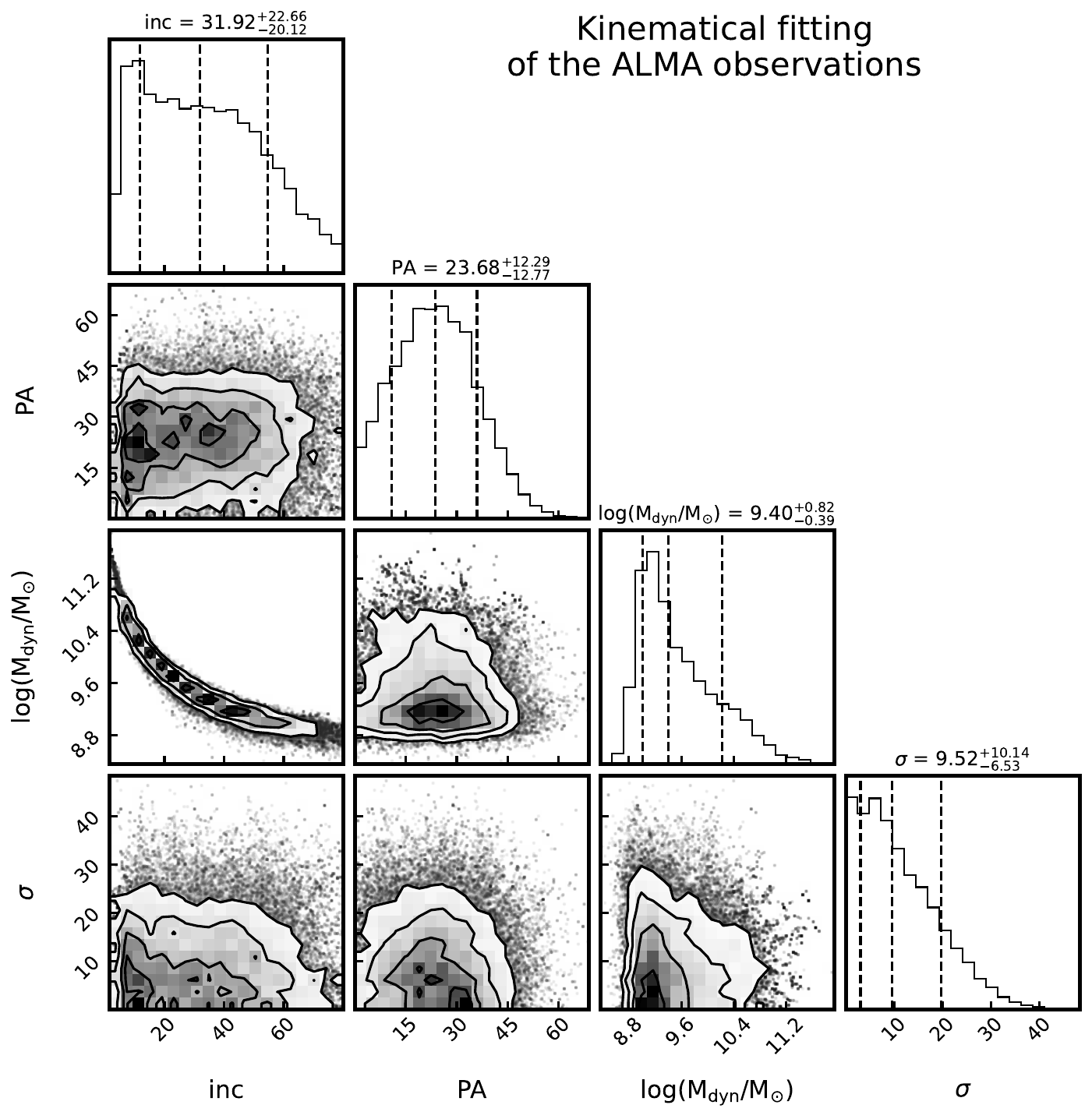}
        \includegraphics[width=0.48\textwidth]{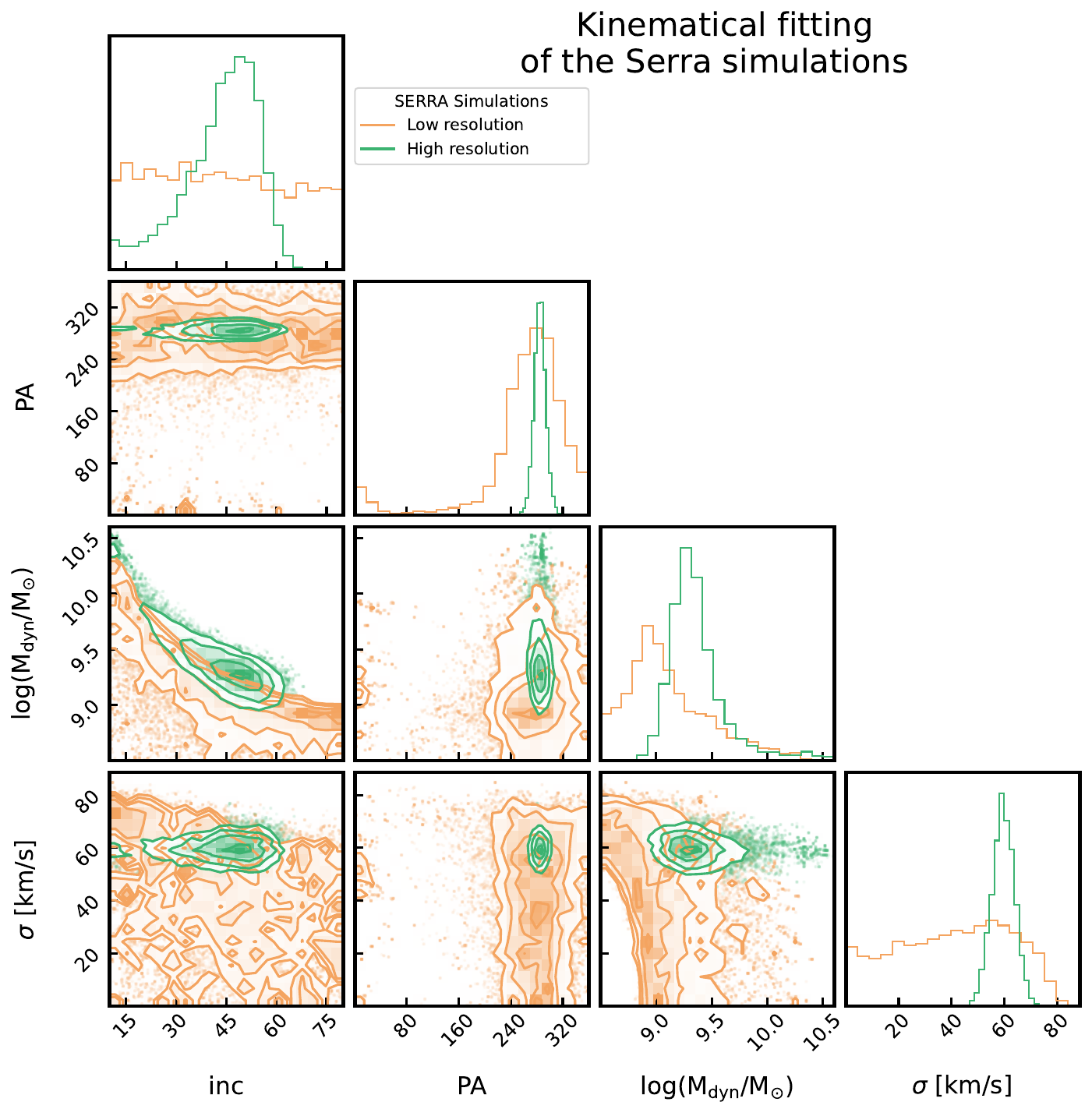}
    \caption{ Posterior distribution from the kinematical modelling of ALMA data of \target (left panel) and  mock observations of the galaxy Amaryllis from SERRA simulations (right panel).}
    \label{fig:corner}
\end{figure*}

We present the posterior distribution of our kinematical modelling of the ALMA data of \target and the mock observations of the galaxy Amaryllis from SERRA cosmological simulations (both low and high resolution) in Fig. \ref{fig:corner}.

\section{ALMA mock observation of Amaryllis during a merger}\label{s.merger}

\begin{figure*}
        \centering
	\includegraphics[width=0.8\paperwidth]{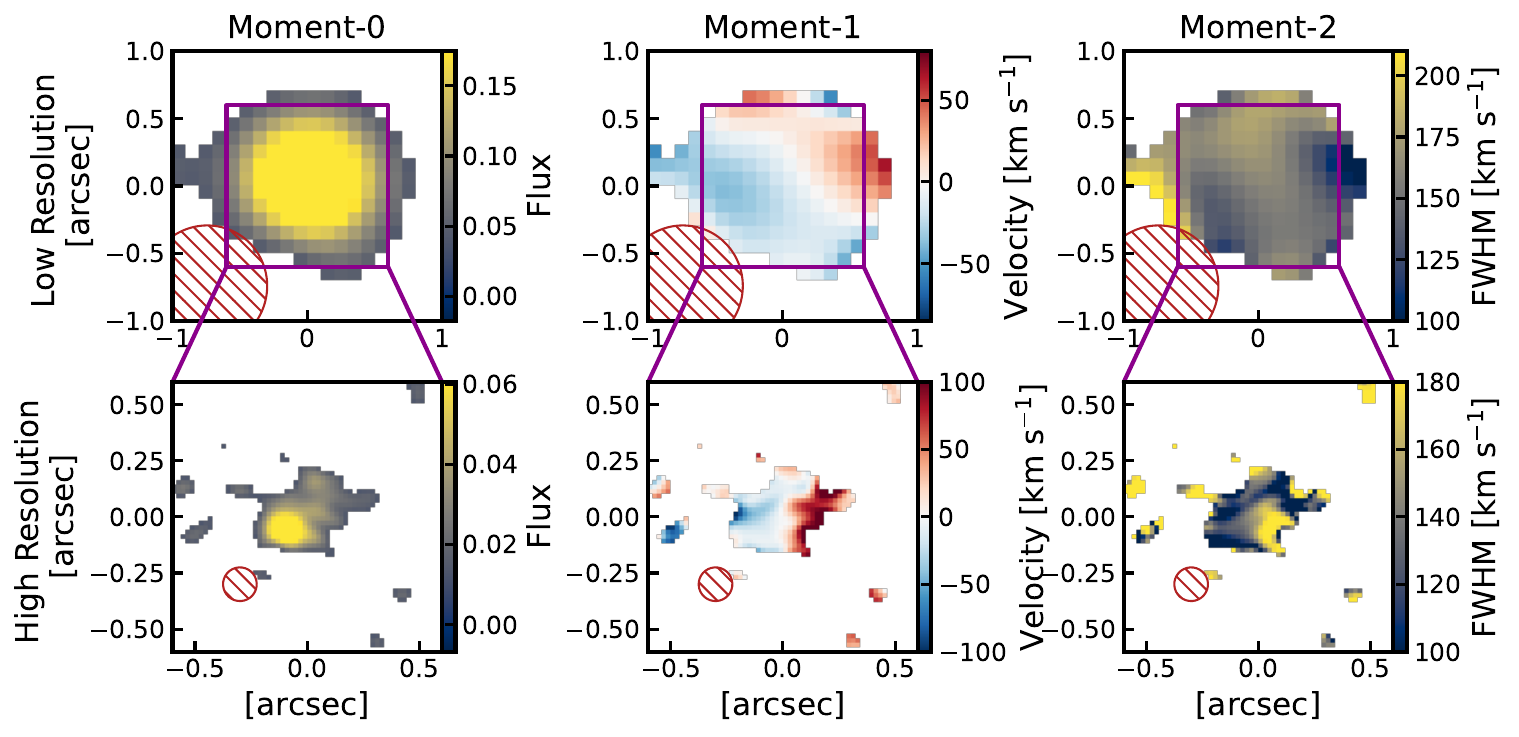}
    \caption{Mock observations of \target analogue in the SERRA simulations - "Amaryllis" during a merger event. The top row shows flux, velocity and FWHM maps of mock low-resolution  ALMA observations (0.9\arcsec), while the bottom row shows mock high-resolution observations (0.15\arcsec). The simulations show that a rotating galaxy shows a small velocity gradient ($<40$ \kms) in the low-resolution data, similar to our observations. The simulations show that a rotating (thin) disc galaxy with a size and mass as \target, and observed under similar conditions, should show a similar velocity gradient to the one obtained.}
    \label{fig:Serra_merger}
\end{figure*}

In Fig. \ref{fig:Serra_merger}, we explore a snapshot of the SERRA simulations where Amaryllis is merging with a close by galaxy. In the low resolution observations (top row) the observations would produce a similar velocity gradient, however such companion would easily be seen in the NIRCam photometry. However, the high resolution ALMA or \jwst data could easily distinguish between a merger and rotating scenario, as the there would be a clearly visible tail in the moment-0 map as well as clearly disturbed velocity field and velocity dispersion map. 

\bibliographystyle{mnras}
\bibliography{mybib}




\appendix


\bsp	
\label{lastpage}
\end{document}